# Robustness of electricity systems with nearly 100% share of renewables: a worst-case study.

Francisco Gutierrez-Garcia, Angel Arcos-Vargas [(*)] & Antonio Gomez-Exposito

**University of Seville, Spain**

**Abstract**

Several research studies have shown that future sustainable electricity systems, mostly based on renewable generation and storage, are feasible with today's technologies and costs. However, recent episodes of extreme weather conditions, probably associated with climate change, cast shades of doubt on whether the resulting generation portfolios are sufficiently robust to assure, at all times, a suitable balance between generation and demand, when adverse conditions are faced. To address this issue, this work elaborates a methodology intended to determine a sustainable electricity system that can endure extreme weather conditions, which are likely to occur. First, using hourly production and demand data from the last decade, along with estimates of new uses of electricity, a worst-case scenario is constructed, including the storage capacity and additional photovoltaic power which are needed to serve the demand on an hourly basis. Next, several key parameters which may have a significant influence on the LCOE are considered, and a sensitivity analysis is carried out to determine their real impact, significance and potential trends. The proposed methodology is then applied to the Spanish system. The results show that, under the hypotheses and conditions considered in this paper, it is possible to design a decarbonized electricity system that, taking advantage of existing sustainable assets, satisfies the long-term needs by providing a reliable supply at an average cost significantly lower than current market prices.

Total: 225 words

*Highlights:*

- A methodology to assess the robustness of renewable electrical systems is developed.
- The LCOE sensitivity to several key parameters is analyzed.
- The required storage capacity is modest when at least 40% of hydro is dispatchable.
- The worst-case LCOE for the Spanish system is lower than current wholesale prices.
- Expected PV and storage cost reductions will further reduce the LCOE obtained

---

[(*)] Camino de los descubrimientos s/n. 41092 Seville (Spain).   aarcos@us.es

# 1 Introduction.

Over the recent years, most countries have established different pathways to massively integrate renewable energy sources (RES) into their electric energy systems, keeping in mind two main goals: to develop a more secure and efficient energy system, allowing the population to obtain cheaper electric supply; and to meet the environmental objectives committed in the Paris Agreement [1]. In particular, Spain is one of those countries which has not yet fully developed its high RES potential. The Spanish renewable policies, reflected in the National Energy and Climate Plan (NECP) for 2021-2030 [2], consider ambitious objectives for the full decarbonization of the energy system, starting with the elimination of the so-called "sun tax" imposed by past governments [3-4] and including an estimation of the required public and private investments. In this volatile context, the definition of credible future renewable scenarios, along with their feasibility and cost assessments, become key issues to be faced during the years to come.

Indeed, the analysis of future renewables scenarios is a hot topic in the literature, which has been growing in the last decade. Many papers evaluate the pathways from very different points of view and criteria, considering varying degrees of renewable shares in the system, establishing different time horizons for RES deployment, diverse geographical spans (regions, countries, etc), focusing on technological, economic, social and/or environmental aspects, etc. Some authors define the initial conditions from the current scenario and perform an analysis of the expected evolution up to the target year, considering certain constraints, while others define their own future scenarios and assess possible pathways to achieve them. Beyond those factors, significant differences can be observed regarding the methodologies used in building and assessing the involved scenarios.

On the one hand, for the assessment of future scenarios, some works adopt methods and algorithms previously developed by other authors or companies. A common model used in several papers [5-8] is the EnergyPLAN [9] developed by Aalborg University (Denmark), for the hourly analysis of regional energy systems over a year. Ref. [5] addresses two renewable scenarios for Macedonia, with a share of 50% and 100% for 2030 and 2050 of RES respectively, combined with pumped hydro storage and electric vehicle (EV) batteries as energy storage systems. Ref. [6] includes variations of wind installed capacity in the model in order to calculate the optimal installed capacity and analyses the influence of wind energy share into the system for two scenarios, including storage systems. Combining the EnergyPLAN model with H2RES [10], a 100% renewable scenario is developed for Croatia in [7], which is aimed at evaluating the role of the storage technology in a renewable self-sufficient energy system, from a technical, economic and environmental perspective. The full transition for the current Portuguese energy system to a renewable system is assessed in [8] for different possible future strategies, duly considering the role of grid interconnections as stabilizing resources that prevent RES curtailment, while also pointing out the importance of energy storage systems. A life-cycle assessment (LCA) is carried out in [11], based on the methodology proposed in [12], to analyse the sustainability of the UK energy system through 36 different (techno-economical, environmental and social) parameters, comparing five future energy mix scenarios with the energy system in 2009, each differing in the share of RES, nuclear and fossil. Additionally, in [13], a modelling framework of an integrated hybrid LCA model is presented, based on [14] and covering nine world region for 2050, for the evaluation if the environmental impact through climate change mitigation scenarios. The results for the European scenario are considered in [15], which assesses the environmental impact of a combination of 44 scenarios, optimizing the total system costs for different shares of low-carbon fossil energy options, variable RES technologies and storage systems.

The Brazilian power generation paradigm change is faced in [16] with a Multi-Criteria Decision Analysis (MCDA), whereby five scenarios for 2050 are evaluated and ranked depending on the perception of experts on the scenario evaluation process. The Business as Usual (BaU) case is analysed as baseline scenario and four alternative scenarios are designed in accordance with [17] and modelled using MESSAGE-Brazil v.1.3 integrated model, originally developed by IAEA [18], with a higher integration of RES. In ref [19], an analysis is performed focused on the RES potential in the energy mix of Thailand and Indonesia from base year 2010 to 2050, considering different renewable energy policy scenarios. The LEAP energy model [20] (developed by Stockholm Environmental Instituted) is used to simulate three scenarios for each country (reference, renewable and renewable potential) with the aim to obtain the $CO_2$ emissions and electric cost of production and to compare the results with the reference case. The variability and uncertainty of wind and solar resources is assessed in [21] regarding the integration of RES in the United States electric systems, focusing on the feasibility and implications, from a technical, economical and sustainable perspective, of six highly renewable scenarios for 2050, using ReEDS model [22]: two baseline scenarios which reflect futures with continued reliance on conventional generation and four scenarios where 80% of all electricity generation in 2050 is sourced from renewables (wind, solar, geothermal, biomass and hydropower). Ref [23] is aimed to establish a roadmap towards 100% RES generation in Nordic countries. Besides considering the role of high level of solar and wind shares in the renewable transition, this study evaluates with the TIMES-VVT algorithm, developed by [24], the feasibility of power-to-gas technology to enable the utilisation of sustainable synthetic gas (methane), produced from 100% renewable electricity, in other sectors beyond power generation.

On the other hand, some researchers develop their own methodologies and algorithms in order to analyse future scenarios and roadmaps. Ref. [25] performs an evaluation of a system composed of 30 interconnected European Countries, and estimates the optimal scenario minimizing the energy sources cost using a constrained linear optimization model in which a sensitive analysis of weather data, cost parameters and policy constraints is performed. In [26], a theoretical model is defined in order to perform an environmental, economic and social evaluation of the potential contribution of PV energy to sustainable development, taking into account the PV deployment roadmaps of several international institutions (International Energy Agency [27, 28], Green Peace [29] and International Renewable Energy Association [30]). The arbitrage potential of wind and solar surplus in a 100% renewable generation in Denmark is analysed in [31], with the help of a linear algorithm called WDRESM (Weather-Driven Renewable Energy System Modeling). This tool considers both renewable generation and storage system, and calculates the required storage capacity, firstly, for the optimal balance and, later, for the optimal storage volume. Furthermore, both [32] and [33] simulate hourly-based scenarios considering the inclusion of RES in the respective regions using their own algorithms. Ref. [32] proposes a new 100% renewable electricity model for the Australian National Electricity Market with the aim to identify and quantify the challenges associated to warranting a reliable system. Ref. [33] considers all the municipalities of Spain and evaluates, with a MATLAB algorithm [34], the rooftop PV potential and the feasibility and economic impact of deploying a brand new system based on rooftop PV and battery storage systems. Moreover, alternative scenarios, comprising the current renewable portfolio and the future RES facilities considered in the Spanish NECP, are analysed, in order to determine least-cost combinations of PV and battery storage systems as a function of hydro generation dispatchability. Likewise, [35] proposes a new specific energy system model for the power generation in Spain, integrating LCA indicators of current and future electricity generation technologies.

Also, in this regard, based on so-called holistic indicator RESI (renewable energy security index) defined in [36], [37] explores the relationship between energy security and climate change by combining LCA and energy system methodology (ESM). Two scenarios for 2050 are analysed (BaU, in accordance with [35], and an alternative energy security scenario) aimed to obtain the mix with optimal energy system costs and to establish an energy policy-making process for achieving that mix. LUT develops a model in [38] for calculating the least-cost energy transition pathways with high levels of spatial (20 regions in Europe) and temporal (hourly) resolution, pointing out the versatility of solar PV and its role in the European energy transition. Several simple logistic curves are used in [39], in combination with the data of the last two decades, for simulating and forecasting the growth and penetration of RES technologies and the increase of the electric energy consumption in India for six different scenarios. By defining an economic model for calculating CAPEX, based on the capacity factor and $CO_2$ emissions, [40] evaluates four potential renewable energy scenarios (BaU, strengthened solar, strengthened wind and suggested scenario) for 2030 in South Korea, aimed at achieving the objective proposed by South Korean Government [41]. The results of each scenario are compared with the BaU case, to assess the influence of the wind and solar installed capacities in the total Korean energy mix, from an economic, environmental and feasibility perspectives.

Finally, some papers carry out a literature review of previous assessments and/or define their own scenarios according to the results described in the reviewed publications. In Ref [42], four different scenarios are proposed and analysed for the future of RES, using the BaU scenario as reference, and focusing the others on the climate change, the energy security and a combination of both criteria for a clean and secure energy future of many countries and world regions. More than 50 papers are reviewed in [43], in order to carry out a LCA along the guidelines provided by ISO 14040 [44] and 14044 [45], in which the performance of the energy production technologies is harmonised according to the results found in the revisions. Ref. [46] is focused on evaluating the potential of power systems to deal with high integration of variable RES in future flexible power systems, considering both generation and storage technologies.

Table 1 summarizes the publications reviewed above, providing information about the model used in the evaluations and the scope of the assessment.

*Table 1. Literature review summary.*

| Reference | Model | Economical assessment? | Environmental assessment? | Extension |
|---|---|---|---|---|
| [5] | EnergyPLAN model | Yes | Yes | Macedonia |
| [6] | EnergyPLAN model | Yes | No | United Kindgom |
| [7] | EnergyPLAN model + H2RES | Yes | Yes | Croatia |
| [8] | EnergyPLAN model | Yes | No | Portugal |
| [11] | LCA | Yes | Yes | United Kingdom |
| [13] | Hybrid LCA | No | Yes | 9 World regions |
| [15] | Hybrid LCA | No | Yes | Europe |
| [16][1] | MCDA | No | No | Brazil |
| [17] | MESSAGE | Yes | Yes | Brazil |
| [19] | LEAP energy model | Yes | Yes | Indonesia and Thailand |
| [21] | ReDs model | Yes | Yes | Unite States of America |
| [23] | TIMES-VVT | Yes | No | Nordic countries (Norway, Denmark, Sweden, Finland) |
| [25] | Own model | Yes | Yes | Europe |
| [26] | WDRESM | Yes | Yes | World |
| [32] | Own model | No | No | Australia |
| [33] | Own model | Yes | No | Spain |
| [35] | Own model and indicators | Yes | Yes | Spain |
| [37] | LCA and own indicators | No | Yes | Spain |
| [38] | Own model | Yes | Yes | Europe |
| [39] | Logistic curves | No | No | India |
| [40] | Own model | Yes | Yes | South Korea |
| [42] | Literature review pooling | Yes | Yes | World |
| [43][2] | LCA | No | Yes | - |

1) Multicriteria analysis of 5 proposed scenarios including costs and CO2 emissions as two parametres among a total of 16.

2) Life cycle assessment performed with data of energy technologies from literature review for harmonising results.

To sum up, a lot of analyses can be found in the literature regarding the assessment of technological, socioeconomic and environmental aspects of RES integration, but very few are focused on evaluating the robustness of the resulting scenarios against the yearly variability of RES. Some of those studies reveal the significant influence that the long-term uncertainty of non-dispatchable RES generation (e.g., wet vs. dry year, windy vs. calm year, etc.) can have on the reliability of energy supply, with the associated risks of not being able to withstand the most adverse combinations of generation and consumption. Indeed, totally unexpected, rare meteorological phenomena, such as the Filomena storm in the Iberian Peninsula [47], or the heavy snowfall in Texas [48], have recently shown that major blackouts and/or socially unacceptable electricity prices can take place if there is a temporary shortage of generation (renewable or conventional) to serve the peak demand. Moreover, such worst-case scenarios recur periodically in different systems. For instance, in the case of Texas, a similar incident took place in 2011 [49], but the measures taken since then have not been sufficiently effective, in view of what has happened ten years later. Consequently, there is a clear need to more deeply study worst-case scenarios that could trigger electric supply failures, before it can be stated whether a whole system can safely meet the electricity needs of the system under study.

In this conceptual framework, given the time scales involved in the system adaptation, we can formulate the following research questions:

1. What should the RES portfolio be, so that it can deal with the worst-case expected scenario?
2. How will the variability of the wind/hydro year affect the resulting LCOE?
3. How will the configuration and costs of the added PV, along with those of the required storage, affect the LCOE?

To answer those and other related questions, this work performs a feasibility analysis of the electrical energy mix that may arise in Spain beyond 2035-40, when RES penetration levels exceeding 85-90% are expected. For this purpose, the hourly electric generation and consumption profiles over the last decade are considered. Nuclear and fossil generation production (except for cogeneration) is eliminated from the hourly balances, and the gap is filled with a least-cost combination of new RES facilities (wind and solar) along with battery storage. After evaluating each year separately, the worst-case scenario is defined by identifying throughout the entire decade the most unfavourable combination of hourly energy production that matches the electric demand. The assessment considers the same technologies and hypotheses as in the sustainable scenario proposed in [33], albeit not so much focused on rooftop PV, and compares the resulting LCOE for every year with that of the worst case. Moreover, the sensitivity of the LCOE with respect to several key parameters is analysed. These include: 1) share of dispatchable generation; 2) time alignment between the RES production and the net demand; 3) utilisation factor of the installed RES capacity, 4) share of PV system installed on the ground vs rooftop, and 5) battery costs.

The paper is organised as follows: Section 2 details the different steps of the proposed methodology, which is applicable to any electricity system for which the required data are available. In Section 3, the assessment method is applied to the Spanish case over the decade 2010-2019, while Section 4 discusses the results obtained, regarding the relevance of every analysed parameter. The paper concludes with further reflections on the addressed research questions.

## 2 Methodology

The methodology used in this study consists of three major steps. Firstly, for every year considered, the least-cost combination of RES and storage capacity satisfying the hourly demand is determined, duly considering the existing and expected RES assets. Secondly, a synthetic yearly scenario is created based on the worst-case combinations of RES production for the decade considered. The worst-case year is assessed in the same way as regular years, and the resulting LCOEs are calculated. Finally, a sensitivity analysis of the LCOE with respect to several key parameters, characterizing the hourly RES production and unitary costs, is performed. Each step is described in detail below.

## 2.1. RES portfolio optimization

The proposed method to determine, for a given year, the optimal combination of RES and storage systems, follows [33]. The salient features of this methodology are summarized as follows (see the Annex):

1) It considers the hourly electricity actually produced by each RES technology (wind, PV solar, thermal solar, hydropower and biomass) in the year considered, while removing the contribution of nuclear and fossil generators (cogeneration assets are retained).

2) An estimation of the energy consumed by light-duty vehicles is performed, assuming the entire fleet is electrified, which is added to the hourly base-case demand.

3) The hourly electricity gap, created by the removal of thermal generation units and the addition of EV demand, is assumed to be filled by a least-cost combination of PV energy and battery energy storage. The hourly contribution of the additional PV is based on the data provided by the Photovoltaic Geographical Information System (PVGIS) [50].

The cost optimization is performed in terms of the LCOE (Levelized Cost of Energy) for the whole system, including the additional PV and storage capacity, defined as follows:

$$LCOE = \frac{1}{D} \cdot \sum_{i=1}^{n} \frac{P_i \cdot I_i}{R_i} \qquad (1)$$

Where $D$ is the annual electricity demand, $n$ is the number of generation technologies considered, $P_i$ is the installed power for technology $i$, $I_i$ is the unitary investment cost per unit power, and $R_i$ is the depreciation period.

Such definition of the LCOE, based on [51], is a simplification of the actual formulation made by the Joint Research Centre of Europe [52], based on two assumptions: 1) the temporal evolution of capital recovery factors is ignored; and 2) the OPEX (operational and maintenance costs) is negligible compared with the CAPEX.

The least-cost scenario is obtained through an iterative exhaustive search process, in which all feasible combinations of additional PV power and storage capacity are sequentially explored and the one with lowest cost is identified. In this context, a feasible solution is the one for which sufficient electricity production (coming from RES or storage) can be committed to cover the demand for every hour of the year. The reader is referred to the Annex for more details.

## 2.2. Worst-case scenario assessment

After performing the hourly energy balance and calculating the resulting LCOE for every year of the last decade, a hypothetical synthetic year, representing the worst-case scenario, is created by combining the most adverse conditions of energy generation and consumption.

Regarding the electricity production, the worst year for each technology is simply the one with the lowest capacity factor, defined by the ratio of the produced energy and installed power. For instance, in Spain, the capacity factor of the wind energy over the last decade (2010-2019), lies within the interval 1900-2400 h, the worst year being 2011. However, the worst year in terms of hydroelectricity was 2017, with 846 h, in contrast to the 2055 h of the best year in the decade (2010).

As for the electricity demand, the worst case is defined in a similar fashion, except for the fact that the four seasons are separately considered. So, the worst year, in terms of demand, is a "frankestein" year composed of the four seasons with largest consumption in the decade considered. This way, we do not exclude the possibility of having to face a year in the future with very cold winter and high temperatures in summer.

The worst case so defined accounts for the lowest contribution of each existing technology and the highest electricity consumption, leading to the largest possible gap between the existing installed portfolio and the expected demand. Then, the algorithm described in Section 2.1 is applied to this synthetic scenario in order to obtain the optimal additional PV power and storage system, and the resulting LCOE estimation.

Figure 1 represents the methodology outlined above, for the assessment of the RES portfolio and definition of the worst-case scenario.

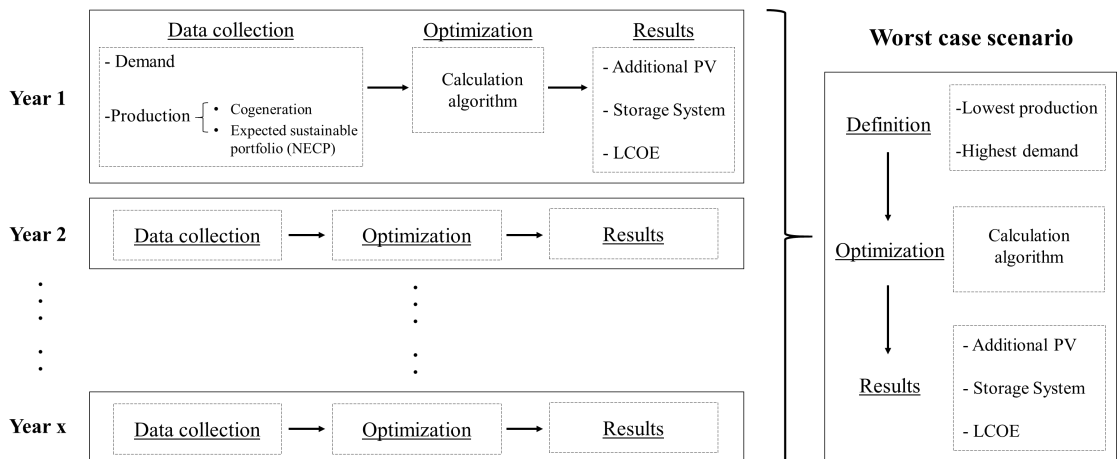

*Figure 1. Overview of the input data, processing blocks and results involved in the proposed assessment.*

## 2.3. Sensitivity assessment

Apart from the installed capacity for each technology, the main outcome of the portfolio optimization process described above is the resulting LCOE for every year considered (in this work a decade plus the worst-case, synthetic year). The next step consists of performing an assessment of the influence that several key factors exert on the costs. Those factors have to do with the RES generation uncertainty and dispatchability, alignment with the demand and expected evolution of costs, as discussed in the sequel:

1) RES dispatchability. In this work, only the hydro generation is assumed to be partly dispatchable in time. Indeed, cogeneration (and biomass) could also shift its production in time to a certain extent, but this possibility will be neglected, as cogeneration production is significantly conditioned by the steam needs of the associated industrial processes. Specifically, in addition to the base case assumption (40% of hydro is dispatchable) additional dispatchability ratios are considered, up to 85%. Then, the portfolio optimization algorithm is run for each dispatchability ratio assumed, and the impact on the resulting LCOE is quantified. Logically, the more dispatchable the RES generation, the lower the required storage capacity and associated LCOE.

2) RES alignment with demand. A coincidence factor (CF) is introduced for each RES technology, characterizing the extent to which its production matches or is synchronized with the residual demand, i.e., the demand not yet served by the remaining RES sources. The CF for technology $i$ is mathematically defined as the scalar (or inner) product of the hourly production vector, $\boldsymbol{E}_i$, and the residual or remaining electric consumption vector, $\overline{\boldsymbol{D}}_i$, both with 8760 components in a regular year (leap years with 8784 components), duly normalized as follows:

$$CF_i = \frac{\boldsymbol{E}_i^T \cdot \overline{\boldsymbol{D}}_i}{||\boldsymbol{E}_i||_2\, ||\overline{\boldsymbol{D}}_i||_2} < 1 \qquad (2)$$

The residual or remaining electric demand, $\overline{\boldsymbol{D}}_i$, is defined by:

$$\overline{\boldsymbol{D}}_i = \begin{bmatrix} D_i(t_1) \\ D_i(t_2) \\ \ldots \\ D_i(t_{8760}) \end{bmatrix}; \qquad \overline{\boldsymbol{D}}_i(t) = \begin{cases} 0, & \text{for } \boldsymbol{D}(t) - \sum_{k \neq i}^{n} \boldsymbol{E}_k(t) \leq 0 \\ \boldsymbol{D}(t) - \sum_{k \neq i}^{n} \boldsymbol{E}_k(t), & \text{otherwise} \end{cases} \qquad (3)$$

where $\boldsymbol{D}(t)$ is the total demand at hour $t$, and $\boldsymbol{E}_k(t)$ the hourly energy produced by the non-dispatchable technology $k$.

3) Net utilization factor (UF): relates the fraction of annual energy produced by a given technology, which is actually used, with its installed power. Mathematically, the net utilization factor for technology $i$ is defined as follows:

$$UF_i = \frac{H_i}{P_i} \qquad (4)$$

Where $H_i$ is the annual energy produced by technology $i$ ($E_i$) minus the curtailed energy, and $P_i$ is the corresponding installed power. Note the difference of the UF defined by (4) with respect to the conventional capacity factor, which assumes that all the energy produced is consumed (no spillage occurs).

Besides the UF for each technology separately, the UF is also computed for the overall electricity generation, which is referred to the total installed power, including the additional PV facilities.

4) Share of rooftop PV. A major unknown in future decarbonized scenarios, particularly in countries such as Spain with low latitudes, is the share of PV facilities that will be deployed on rooftops, with unitary costs clearly exceeding those of utility-scale facilities on ground. So, in addition to the base case scenario, where all additional PV is assumed to be built in the form of large plants, the LCOE is recalculated and compared for increasing shares of rooftop PV, in order to assess the impact of having more decentralized RES, which do not require additional land surfaces.

5) Storage costs. Finally, the influence on the resulting LCOE of the cost of massive storage systems is assessed. For this purpose, a sensitivity analysis is performed for several possible values of battery costs, that will fluctuate depending on many factors, such as policies, maturity of competing technologies (Li-ion, NaS, flow batteries), availability of raw materials, etc. This way, a range of LCOE values is obtained, representative of future foreseeable conditions affecting the costs of battery storage.

## 3 Case Study

In order to apply the methodology proposed above to the Spanish case, the hourly data corresponding to generation, demand and costs for the last decade are first gathered, from which the least-cost RES plus storage portfolios are obtained for each year and for the worst-case scenario. Then, a sensitivity analysis is carried out for the relevant factors described above.

### 3.1. Base case scenario

According to the proposed methodology, the necessary data and parameters to define the base case are listed below:

- Hourly generation over the last decade (summarized in Table 2).
- Hourly demand over the last decade (summarized in Figure 4).
- % of hydro generation dispatchability = 40%.
- % ground additional PV = 100%.
- Cost of kWp PV = 500 €.
- Cost of kWh of storage (Li-Ion): 100 €.

The data collection comprises two major components: data corresponding to the energy production and consumption during the last decade, and economic parameters required for the LCOE calculation.

Regarding the electricity production, the Spanish Transmission System Operator (REE) daily provides the contribution of each generation technology to the electricity mix, duly separating the data corresponding to the islands from those of the continental Spain [53]. Each daily report, in our case spanning the decade 2010-19, contains power datasets, which are equivalent to hourly energy values.

The hourly electricity consumption is estimated by combining the contribution of all technologies (including non-renewable sources) with the electricity imported from or exported to the neighbour countries and Balearic Islands, through the existing interconnections. Table 2 provides the Spanish electricity mix for the last decade.

*Table 2. Spanish electricity mix (2010-2019).*

| Year | Nuclear | Fuel/Gas | Coal | Combined cycle | Wind | Hydro | PV | Thermal solar | Thermal renew.(1) | Cogen. & waste | Int. exchange | Balearic link | Demand |
|------|---------|----------|------|----------------|------|-------|----|----|----|----|----|----|----|
|      | (GWh)   | (GWh)    | (GWh)| (GWh)          | (GWh)| (GWh) | (GWh)| (GWh) | (GWh) | (GWh) | (GWh) | (GWh) | (GWh) |
| 2019 | 55,957 | 0 | 11,090 | 51,569 | 52,383 | 23,357 | 9,215 | 5,199 | 3,650 | 32,380 | 6,693 | -1,664 | 249,828 |
| 2018 | 53,268 | 0 | 35,434 | 26,910 | 48,926 | 32,648 | 7,578 | 4,441 | 3,596 | 31,928 | 10,965 | -1,215 | 254,480 |
| 2017 | 55,599 | 0 | 42,744 | 34,150 | 47,144 | 17,221 | 7,814 | 5,282 | 3,684 | 31,177 | 8,994 | -1,160 | 252,649 |
| 2016 | 55,687 | 0 | 35,220 | 25,896 | 47,292 | 33,899 | 7,248 | 5,059 | 3,634 | 28,792 | 7,576 | -1,226 | 249,077 |
| 2015 | 54,718 | 0 | 51,350 | 25,850 | 48,006 | 26,004 | 7,839 | 5,085 | 4,615 | 26,962 | -133 | -1,336 | 248,959 |
| 2014 | 54,753 | 0 | 41,575 | 21,579 | 51,205 | 36,372 | 7,794 | 4,959 | 4,781 | 25,596 | -3,554 | -1,281 | 243,780 |
| 2013 | 54,277 | 0 | 37,545 | 24,618 | 54,629 | 34,663 | 7,915 | 4,442 | 5,064 | 31,989 | -6,732 | -1,269 | 247,141 |
| 2012 | 58,531 | 0 | 49,994 | 38,237 | 48,114 | 18,530 | 7,803 | 3,443 | 4,736 | 33,716 | -11,200 | -570 | 251,334 |
| 2011 | 55,039 | 0 | 39,404 | 50,596 | 41,669 | 29,405 | 7,081 | 1,823 | 3,792 | 32,037 | -6,090 | - | 254,755 |
| 2010 | 59,064 | 1,825 | 20,817 | 64,544 | 42,774 | 39,327 | 6,140 | 692 | 3,172 | 30,845 | -8,333 | - | 260,867 |

(1) Thermal renewable considers biogas, biomass and geothermal energy.

In addition to the renewable energy produced by the existing assets, the analysis considers the sustainable power facilities forecasted in the target scenario defined by the Spanish NECP for 2030 [2]. Table 3 shows the installed power capacity over the last decade, along with the provisions made in the NECP. In this work, a hypothetical scenario, well beyond 2030, is considered, in which both nuclear power and combined cycles have been decommissioned.

*Table 3: Evolution of Spanish power capacity and expected values for 2030.*

| Year | Nuclear | Fuel/Gas | Coal | Combined cycle | Wind | Hydro | PV | Thermal solar | Thermal renew.(1) | Cogen. & waste |
|------|---------|----------|------|----------------|------|-------|----|----|----|----|
|      | (MW) | (MW) | (MW) | (MW) | (MW) | (MW) | (MW) | (MW) | (MW) | (MW) |
| 2019 | 7,126 | 0 | 9,222 | 24,562 | 25,257 | 20,412 | 8,594 | 2,306 | 1,071 | 6,239 |
| 2018 | 7,117 | 0 | 9,562 | 24,562 | 23,091 | 20,376 | 4,466 | 2,304 | 859 | 6,305 |
| 2017 | 7,117 | 0 | 9,536 | 24,948 | 22,922 | 20,359 | 4,439 | 2,304 | 852 | 6,400 |
| 2016 | 7,573 | 0 | 9,536 | 24,948 | 22,900 | 20,359 | 4,430 | 2,299 | 743 | 7,277 |
| 2015 | 7,573 | 0 | 10,468 | 24,948 | 22,864 | 20,359 | 4,420 | 2,300 | 742 | 7,361 |
| 2014 | 7,866 | 520 | 10,972 | 25,348 | 22,845 | 19,443 | 4,428 | 2,300 | 1,012 | 7,075 |
| 2013 | 7,866 | 520 | 11,131 | 25,353 | 22,854 | 19,437 | 4,422 | 2,300 | 975 | 7,089 |
| 2012 | 7,853 | 520 | 11,248 | 25,340 | 22,573 | 19,379 | 4,298 | 2,000 | 953 | 7,240 |
| 2011 | 7,777 | 1,492 | 11,700 | 25,235 | 21,091 | 19,156 | 4,047 | 1,049 | 858 | 7,282 |
| 2010 | 7,777 | 2,860 | 11,380 | 25,235 | 20,057 | 19,139 | 3,458 | 682 | 1,050 | 6,992 |
| NECP | 3,181 | 0 | 0 | 24,562 | 50,333 | 24,133 | 39,181 | 7,303 | 1,649 | 4,011 |

(1) Thermal renewable considers biogas, biomass and geothermal energy.

The next step consists of scaling up (in the case of RES) or down (in the case of cogeneration) the actual generation profile for every year, in order to obtain the hourly production for each technology according to the forecasted sustainable scenario of the NECP. However, as the hourly values of the PV production are only available for the period 2016-2018 in [53], considering that the yearly variability of the solar resource is moderate, an average of existing hourly data is taken for the remaining years.

Table 4 shows the hypothetical generation of the RES technologies, on a yearly basis, in accordance with the expected installed power in 2030, along with the total electric demand (actual plus EV) and the net non-served demand, originated by the removal of nuclear units and combined cycles from the generation portfolio. For convenience, the contribution of thermal renewable, cogeneration and waste technologies has been gathered in the same column of the table. In this work, the resulting demand gap is to be filled exclusively by additional PV power, according to the current cost trends and the interest shown by the stakeholders when asking the TSO for new connection points to the transmission system.

*Table 4. Expected RES production for the power capacities forecasted in the Spanish NECP.*

| Year | Wind (GWh) | PV (GWh) | Thermal solar (GWh) | Hydro (GWh) | Thermal renew., cogen. & waste (GWh) | Total Demand (GWh) | Unserved Demand (GWh) |
|---|---|---|---|---|---|---|---|
| 2019 | 104,390 | 66,520 | 16,038 | 27,615 | 36,030 | 310,603 | 60,009 |
| 2018 | 106,648 | 66,486 | 14,075 | 38,668 | 35,524 | 315,255 | 53,854 |
| 2017 | 103,521 | 68,968 | 16,742 | 20,416 | 34,861 | 313,423 | 68,915 |
| 2016 | 103,945 | 64,106 | 16,070 | 40,183 | 32,426 | 309,851 | 53,122 |
| 2015 | 105,680 | 66,855 | 15,936 | 30,827 | 31,577 | 309,733 | 58,858 |
| 2014 | 112,818 | 66,796 | 15,810 | 45,147 | 30,377 | 304,554 | 33,606 |
| 2013 | 120,314 | 66,928 | 15,293 | 43,039 | 37,053 | 307,916 | 25,290 |
| 2012 | 107,285 | 67,026 | 14,944 | 23,076 | 38,452 | 312,108 | 61,326 |
| 2011 | 99,441 | 66,730 | 15,387 | 37,044 | 35,829 | 315,529 | 61,098 |
| 2010 | 107,341 | 66,789 | 15,052 | 49,590 | 34,017 | 321,641 | 48,852 |

The variability of the annual energy production can be observed in Fig. 2, where the maximum, minimum and average yearly values are represented for all the technologies considered in this model. As can be seen, the most volatile source is hydro (36 TWh ± 43 %), followed by wind (107 TWh ± 12%). On the other hand, the standard deviation of PV production for the analysed decade is less than 3.9%.

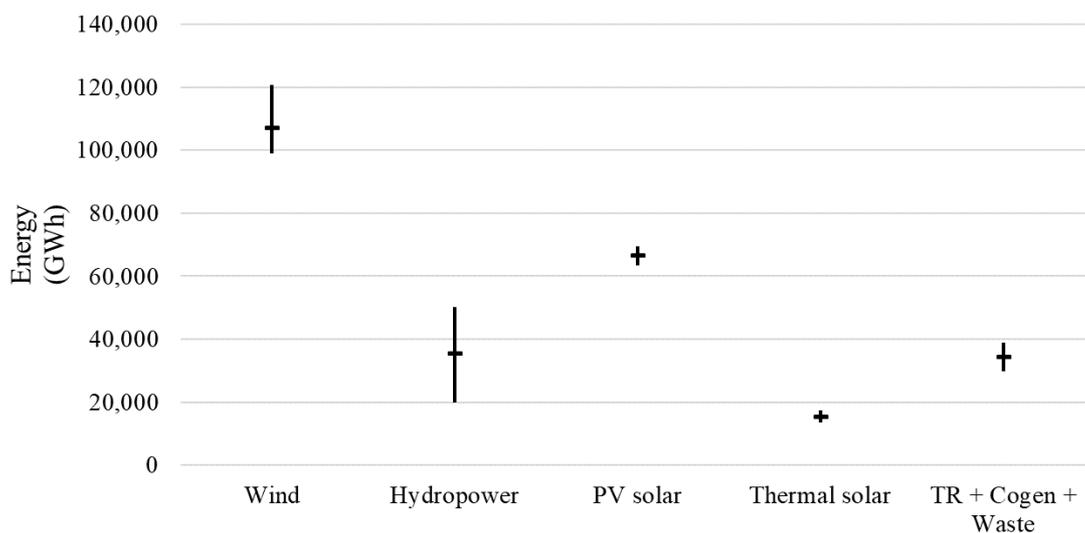

*Figure 2. Maximum, minimum and average RES energy production.*

In this work, the demand of EVs is added to the current demand. For this purpose, according to [33], it is assumed that the Spanish fleet of light-duty vehicles, when fully electrified, would consume about 60 TWh/year. The hourly charging profile for the EV demand is taken from [54], which regularly monitors a representative sample of charging points. The total electric consumption for the decade considered in this work is represented in Fig. 3.

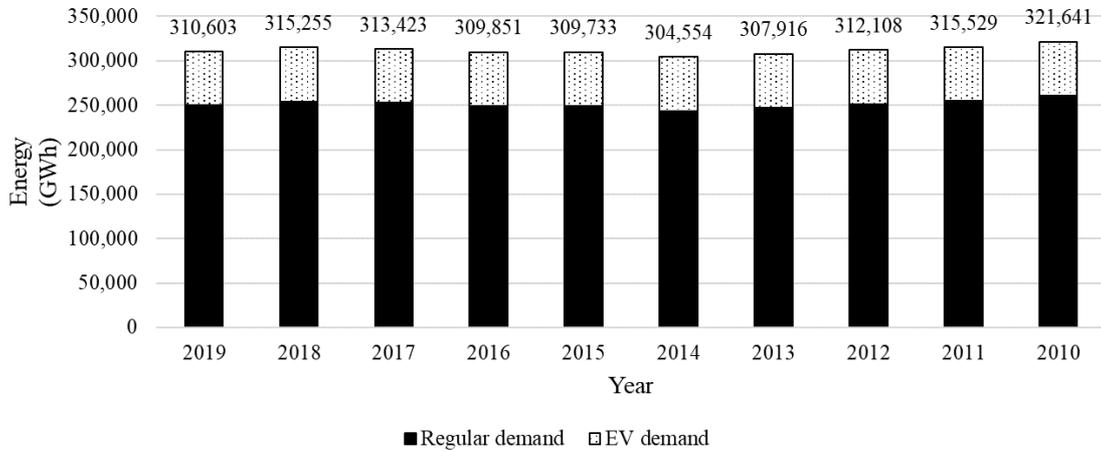

Figure 3. Yearly electricity demand including EVs.

Figure 4 shows the variability of the yearly electricity demand: maximum, minimum and average values for the whole decade, divided by seasons. As expected, winter and summer are the most volatile seasons.

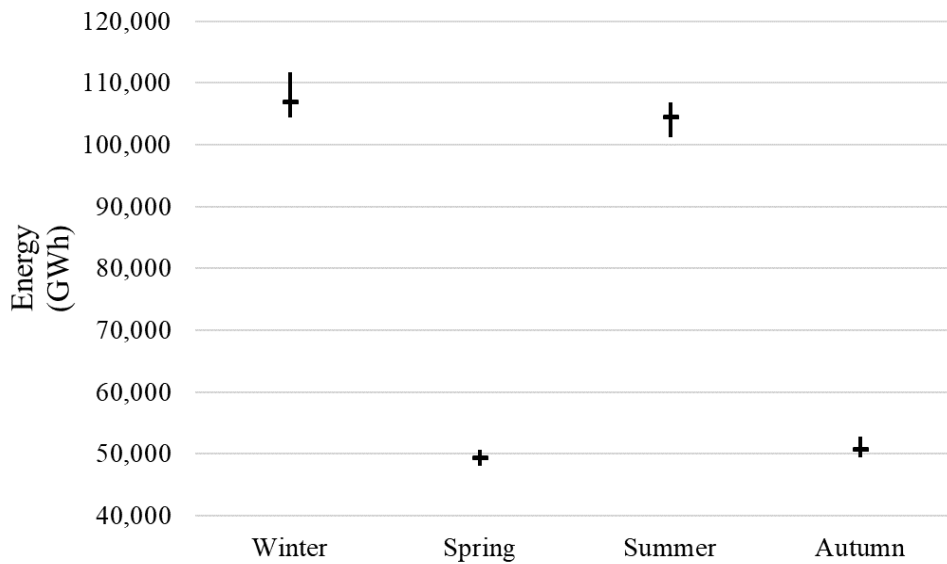

Figure 4: Maximum, minimum and average seasonal electric consumption for the decade 2010-19.

Regarding the second component of the data collection process, Table 5 shows the economic parameters assumed for the estimation of LCOEs: investment cost per unit power and depreciation period for each technology, according to [55], [56] and own estimation. An adaptation of Eq. (1) is performed to duly consider the operational and fuel costs of cogeneration, waste and thermal renewable energy. The contribution of these sources to the LCOE is estimated depending on the used energy and the average market energy price.

*Table 5. Economic parameters involved in the estimation of LCOEs.*

| Technology | Depreciation period (years) | Investment (€/kW) | Energy price (€/MWh) |
|---|---|---|---|
| Actual wind | 25 | 1,300 | - |
| NECP wind | 25 | 900 | - |
| Actual PV | 25 | 1,500 | - |
| NECP PV | 25 | 500 | - |
| Utility scale PV | 25 | 500 | - |
| Thermal solar | 40 | 5,000 | - |
| NECP thermal solar | 40 | 5,000 | - |
| Hydropower | 60 | 2,000 | - |
| NECP hydropower | 60 | 2,000 | - |
| Battery | 13.7 | 100* | - |
| Thermal renewable + cogen + waste | - | - | 50 |

* Investment cost for batteries in €/kWh. This value is a long-term estimation, based on reported trends which consider expected technological advancements and the market structure. However, the impact of this cost on the total LCOE is evaluated in Section 3.4.5., where a range of values for the battery investment costs are considered.

## 3.2. Worst case scenario

By combining the data of energy production and consumption over the last decade, the worst-case hourly scenario is obtained as follows:

1. For each RES technology, the profile corresponding to the year with lowest capacity factor is selected, and then the hourly values are duly scaled to consider the installed capacity foreseen in the NECP. Table 6 presents the values of the capacity factor for each technology and year (lowest marked in red).

*Table 6. Capacity factor (in hours).*

| Year | Capacity factor (h) | | | | |
|---|---|---|---|---|---|
| | **Wind** | **Hydro** | **PV** | **Thermal solar** | **Thermal renew., cogen & waste**[(1)] |
| 2019 | 2,074 | 1,144 | 1,072† | 2,255 | 4,929 |
| 2018 | 2,119 | 1,602 | 1,697 | 1,927 | 4,959 |
| 2017 | 2,057 | 846 | 1,760 | 2,292 | 4,807 |
| 2016 | 2,065 | 1,665 | 1,636 | 2,200 | 4,043 |
| 2015 | 2,100 | 1,277 | 1,774 | 2,211 | 3,897 |
| 2014 | 2,241 | 1,871 | 1,760 | 2,156 | 3,756 |
| 2013 | 2,390 | 1,783 | 1,790 | 1,931 | 4,595 |
| 2012 | 2,131 | 956 | 1,815 | 1,722† | 4,693 |
| 2011 | 1,976 | 1,535 | 1,750 | 1,738† | 4,402 |
| 2010 | 2,133 | 2,055 | 1,776 | 1,015† | 4,230 |

[(1)] Thermal renewable, cogeneration and waste have been considered as a single aggregated source, as REE does not provide separate hourly generation profiles for each of those sources.

† Some values are not totally meaningful, due to high variation in installed power over that year, as can be seen in table 2.

2. The season in the decade with highest demand is used to define the demand of that season in the worst-case year, and then the 60 TWh consumed by the EV fleet are added throughout the year according to the recorded hourly profiles.

To summarize, Table 7 shows the electricity produced by each source and the electric consumption in the worst-case scenario, along with the corresponding year.

*Table 7. Annual electricity balance and gap in the worst-case scenario.*

| Scenario | Wind (GWh) | Hydro (GWh) | PV (GWh) | Thermal solar (GWh) | Thermal renew., cogen & waste[(1)] (GWh) | Demand (Actual + EV) (GWh) | Unserved demand (GWh) |
|---|---|---|---|---|---|---|---|
| Worst Case | 99,441 | 20,416 | 64,106 | 14,075 | 30,377 | 321,921 | 93,506 |
| Reference year | 2011 | 2017 | 2016 | 2018 | 2014 | Winter - 2010<br>Spring - 2010<br>Summer - 2017<br>Autumn - 2010 | |

[(1)] Thermal renewable, cogeneration and waste have been considered as a single aggregated source, as explained in Table 6.

## 3.3 Filling the gap.

In this section, the methodology described in Section 2.1, aimed at filling the gap between the forecasted demand and the RES production forecasted by the NECP, with a least-cost combination of additional PV and storage, is applied to the Spanish case. In the base case scenario, it is assumed that a predefined fraction of 40% of hydroelectricity is fully dispatchable. This yields nearly a 100% renewable energy system (actually 93%, as cogeneration is mostly non-renewable), capable of supplying the required electricity on an hourly basis, under the most adverse conditions found in the last decade.

The algorithm (described in the Annex) provides the optimal amount of PV and battery storage that should be added to the electricity mix, and the resulting LCOE for each year and the worst case, as shown in Table 8.

*Table 8. Additional PV power required (excluding that forecasted by the NECP for 2030), storage capacity and resulting LCOE.*

| Year | PV power (GW) | Storage (GWh) | LCOE (€/MWh) |
|---|---|---|---|
| 2019 | 81.1 | 291.0 | 35.9 |
| 2018 | 74.4 | 272.0 | 34.2 |
| 2017 | 99.7 | 319.0 | 36.8 |
| 2016 | 94.3 | 283.0 | 35.2 |
| 2015 | 93.0 | 300.0 | 35.4 |
| 2014 | 51.8 | 235.0 | **31.9** |
| 2013 | 45.2 | 223.0 | 32.3 |
| 2012 | 98.3 | 312.0 | **37.3** |
| 2011 | 83.7 | 290.0 | 35.1 |
| 2010 | 65.1 | 256.0 | 32.1 |
| **Worst Case** | **146.2** | **370.0** | **38.6** |

As can be seen in the table, the LCOE varies between 31.9 €/MWh and 37.3 €/MWh for the regular years, reaching a maximum of 38.6 €/MWh in the worst-case scenario. It is worth noting that, when adding about 146 GW of PV power, along with just 370 GWh of battery storage, to the RES portfolio forecasted in the Spanish NECP (already assuming 39 GW of PV in 2030), the resulting sustainable system can satisfy the hourly electricity demand under the most adverse conditions found in the Spanish case during the last decade.

Figure 5 shows the energy produced by each technology over the decade along with the electrical demand and the curtailed energy.

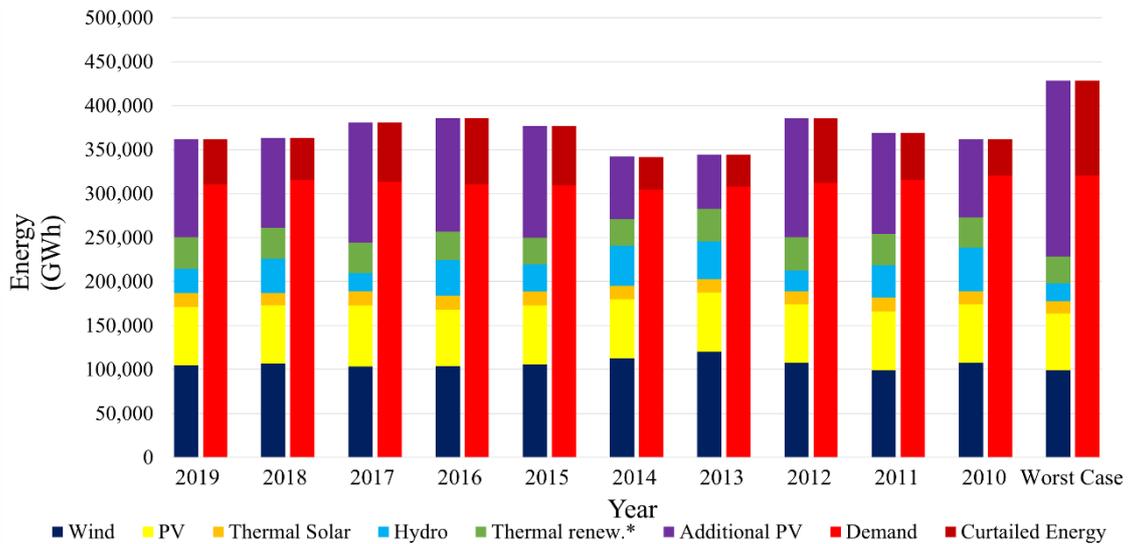

*Figure 5. Electricity share in the base-case sustainable scenarios.*

The curtailed energy varies between 37 TWh (10.1% of the gross demand) in 2013 and 76 TWh (20.9%) in 2016, reaching 108 TWh (31.3%) in the worst-case scenario. This excess of RES energy could be used for alternative applications (e.g., green hydrogen or ammonia production, water desalinization, irrigation of crops, ...), or exported to other countries.

### 3.4 Sensitivity analysis.

Once the base-case scenario is evaluated, the sensitivity of the results against the parameters defined in Section 2.3 is analysed.

#### 3.4.1 Dispatchability.

The RES dispatchability, in this work restricted only to the hydro sources, is considered the single most relevant variable. In order to duly assess the role of this important factor, every year in the last decade, along with the worst-case year, is repeatedly analysed for increasing dispatchability levels of the hydro resource, namely 55%, 70% and 85%, and the resulting LCOE is computed. The application of the optimization algorithm provides the results summarized in Table 9, for each simulated scenario, including the additional PV power, the storage capacity and the LCOE.

*Table 9. Additional utility-scale PV, storage and LCOE.*

| Year | Dispatchability 40% | | | Dispatchability 55% | | | Dispatchability 70% | | | Dispatchability 85% | | |
|---|---|---|---|---|---|---|---|---|---|---|---|---|
| | PV power (GW) | Storage (GWh) | LCOE (€/MWh) | PV power (GW) | Storage (GWh) | LCOE (€/MWh) | PV power (GW) | Storage (GWh) | LCOE (€/MWh) | PV power (GW) | Storage (GWh) | LCOE (€/MWh) |
| 2019 | 81.1 | 291.0 | 35.9 | 75.7 | 277.0 | 35.3 | 71.8 | 266.0 | 34.9 | 69.1 | 256.0 | 34.5 |
| 2018 | 74.4 | 272.0 | 34.2 | 67.8 | 255.0 | 33.5 | 62.5 | 244.0 | 33.0 | 61.1 | 228.0 | 32.5 |
| 2017 | 99.7 | 319.0 | 36.8 | 93.0 | 306.0 | 36.2 | 87.7 | 296.0 | 35.7 | 81.1 | 294.0 | 35.4 |
| 2016 | 94.3 | 283.0 | 35.2 | 79.7 | 273.0 | 34.3 | 71.8 | 257.0 | 33.5 | 67.8 | 240.0 | 32.9 |
| 2015 | 93.0 | 300.0 | 35.4 | 82.4 | 287.0 | 34.6 | 77.1 | 271.0 | 33.9 | 71.8 | 262.0 | 33.5 |
| 2014 | 51.8 | 235.0 | 31.9 | 46.5 | 217.0 | 31.2 | 43.9 | 201.0 | 30.7 | 42.5 | 187.0 | **30.3** |
| 2013 | 45.2 | 223.0 | 32.3 | 39.9 | 203.0 | 31.5 | 37.2 | 185.0 | 31.1 | 34.5 | 174.0 | 30.7 |
| 2012 | 98.3 | 312.0 | **37.3** | 87.7 | 298.0 | 36.5 | 79.7 | 290.0 | 36.0 | 77.1 | 275.0 | 35.5 |
| 2011 | 83.7 | 290.0 | 35.1 | 75.7 | 275.0 | 34.4 | 70.4 | 261.0 | 33.8 | 67.8 | 247.0 | 33.4 |
| 2010 | 65.1 | 256.0 | 32.1 | 59.8 | 237.0 | 31.4 | 55.8 | 222.0 | 30.9 | 54.5 | 205.0 | 30.4 |
| Worst | 146.2 | 370.0 | 38.6 | 131.6 | 360.0 | 37.7 | 123.6 | 345.0 | 37.0 | 117.0 | 334.0 | 36.4 |

The evolution of the LCOE for increasing dispatchability levels is represented in Fig 6. The trend clearly shows how the value of the LCOE decreases as the hydro resource is more dispatchable, owing to the availability of higher amounts of energy to be shifted in time for meeting the demand peaks, and hence the lower storage capacity requirements. Higher dispatchability levels allow the seasonal component of the storage system, which is only useful to meet the electricity consumption in a few peak hours, to be reduced.

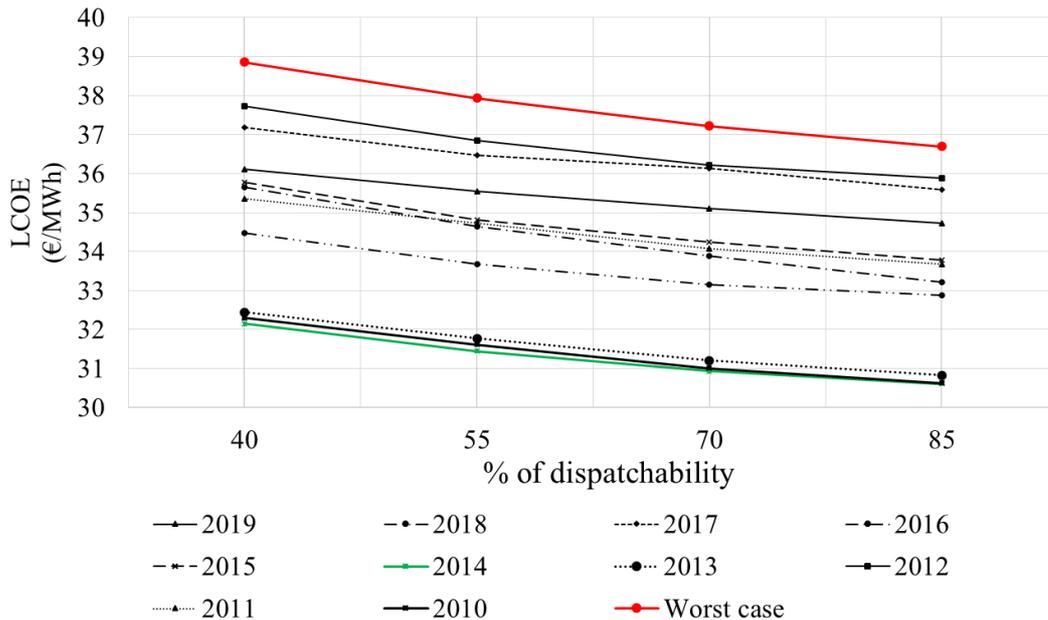

*Figure 6. Evolution of LCOE versus share of dispatchable hydro.*

### 3.4.2 Coincidence Factor

As defined in Section 2.3, for a given technology, the coincidence factor constitutes an indicator of synchronization or alignment between its production and the net demand, throughout the hours of a year. In this section, the variability of wind and hydropower coincidence factors is analysed, and its influence on the LCOE of future foreseeable scenarios is assessed. The analysis of the PV production has been excluded, as the corresponding data is only available for three years, as explained in Section 3.1. Figure 7 shows the results corresponding only to the 40% hydro dispatchability of the base-case (in this and the following figures, the red dot represents the worst-case synthetic year).

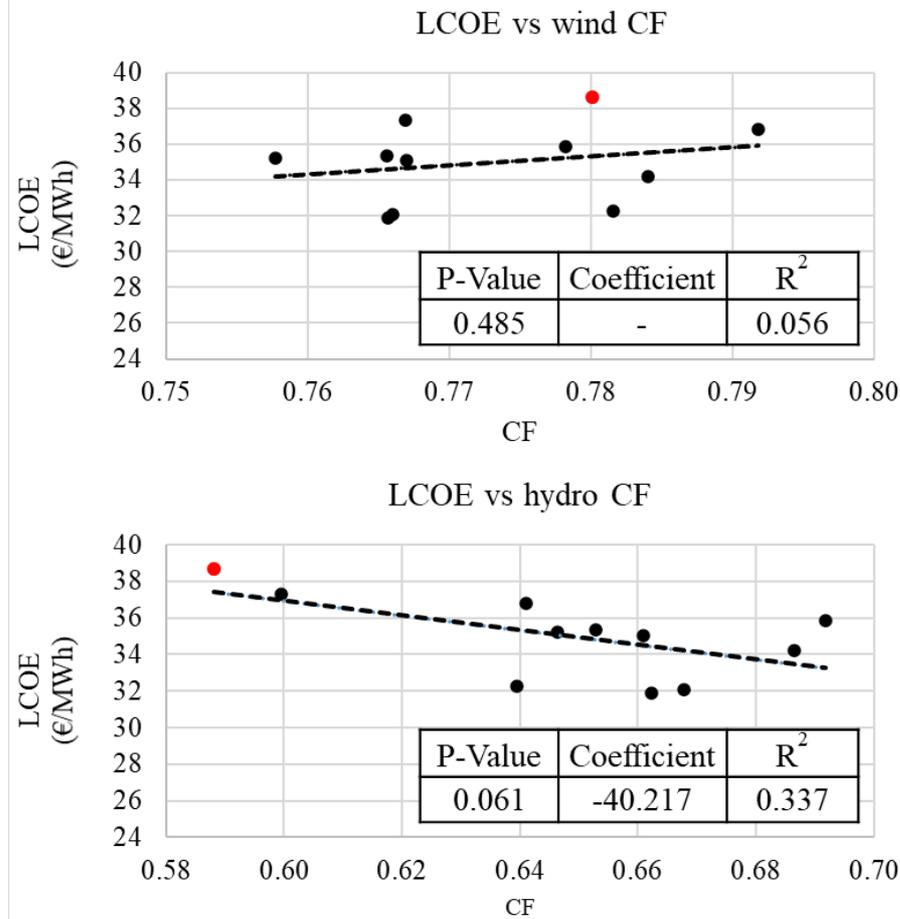

Figure 7 Technology coincidence factor vs LCOE

The high p-value for the wind CF suggests that this variable is not significant in order to explain the LCOE value, which is the reason why it will not be further pursued in this study. In the case of the hydro CF, the significance level is nearly 90%, which means that the null hypothesis cannot be rejected, and so it makes sense to consider the influence of this parameter in the analysis. As might be expected, there is a significant inverse relationship between the hydro CF and the LCOE, the coefficient being –40.2.

### 3.4.3. Utilization Factor

The utilization factor (UF), calculated according to Eq. (4), plays the role of evaluating the annual use of the installed power capacity, which can hence potentially impact the LCOE. The PV technology has been also excluded from this analysis as explained above. Figure 8 shows the results obtained in the evaluation of the UF.

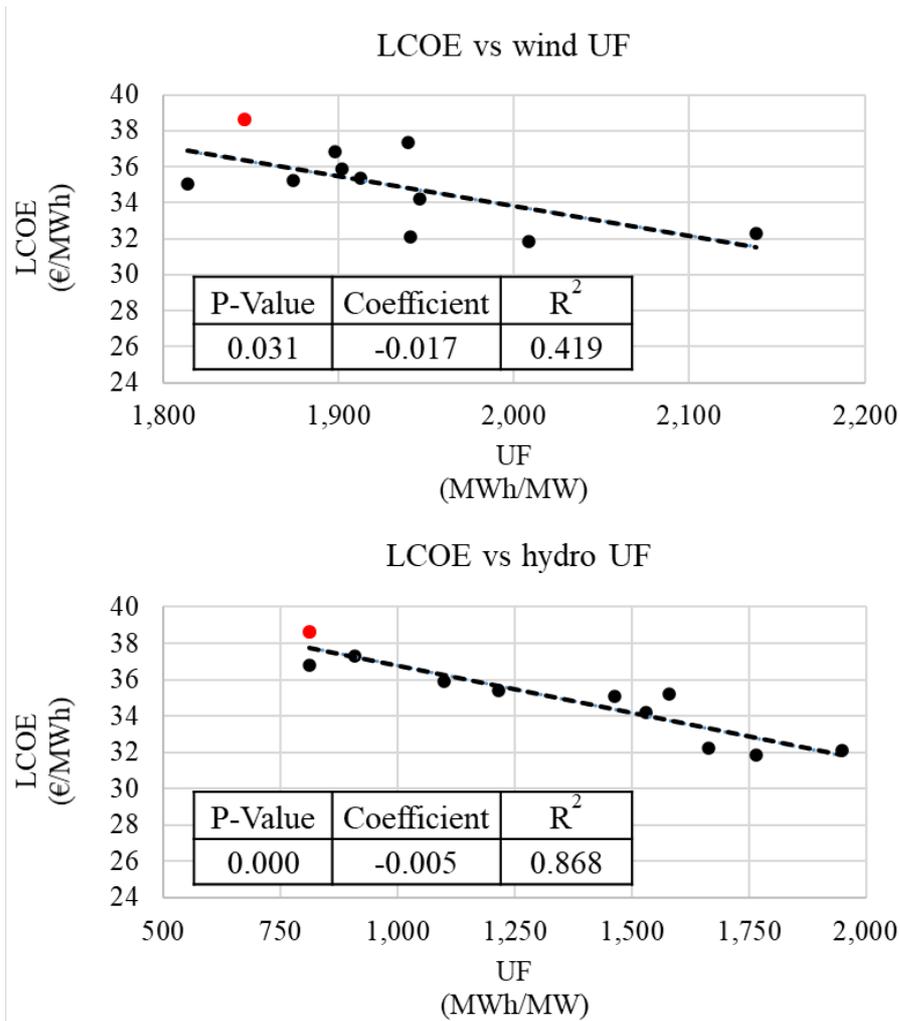

*Figure 8. Utilization factor of PV and wind vs LCOE.*

As can be observed in Fig. 8, both variables are significant (p-value less than 0.05), so the null hypothesis cannot be rejected and therefore their inclusion in the analysis makes sense. The sign of the regressors indicates the existence of an inverse relationship between the values of the UF and the LCOE, as would be expected.

Additionally, the UF of the overall energy production with respect to the total installed power is estimated, in order to assess the correlation between the use of the total capacity and the LCOE. Figure 9 shows the values of the UF factor when applied to the whole portfolio.

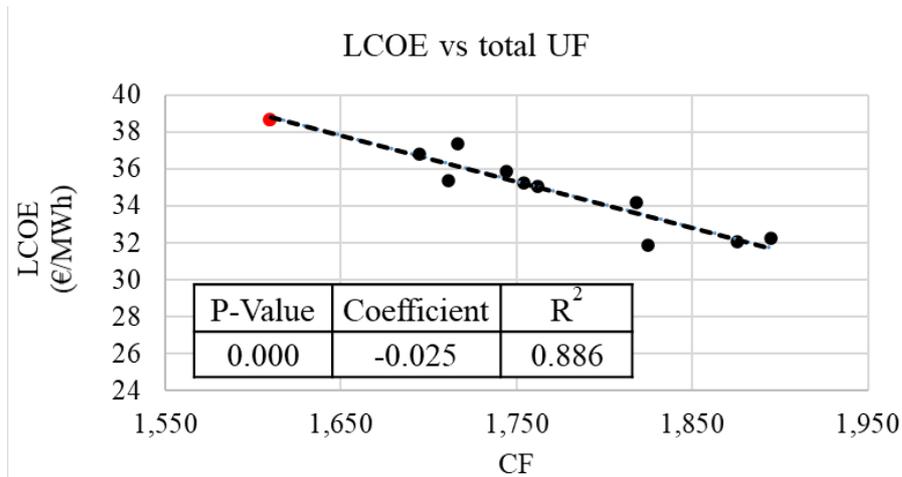

*Figure 9. Total utilization factor vs LCOE.*

It is evident that the UF, when applied to the entire energy production, provides also good results as explanatory variable with a P-value lower than 0.05 ($R^2$ close to 0.9) showing a significant correlation with the LCOE. This is so because, being the installed capacity and the electric consumption approximately the same every year, the net demand to be matched by additional PV and battery storage is lower when the installed power is better harnessed.

### 3.4.4. PV power costs.

In this section, the LCOE for the entire system (average MWh consumed) is analysed for several values of PV investment cost, in order to assess the differences in costs of deploying utility-scale PV (base case scenarios) or rooftop PV.

In addition to that of the base case (500 €/kW), the methodology is applied for three increasing unitary costs, namely 667 €/kW, 833 €/kW and 1000 €/kW, approximately representing hypothetical scenarios in which the share of additional rooftop PV power is 33%, 67% and 100%, respectively, instead of being located entirely on the ground.

The results provided by these simulated scenarios are shown in Figure 10 for the year 2014, considered as an average year, separating the contribution to the total LCOE made by the renewable assets already considered by the Spanish NECP, the additional PV and the battery storage.

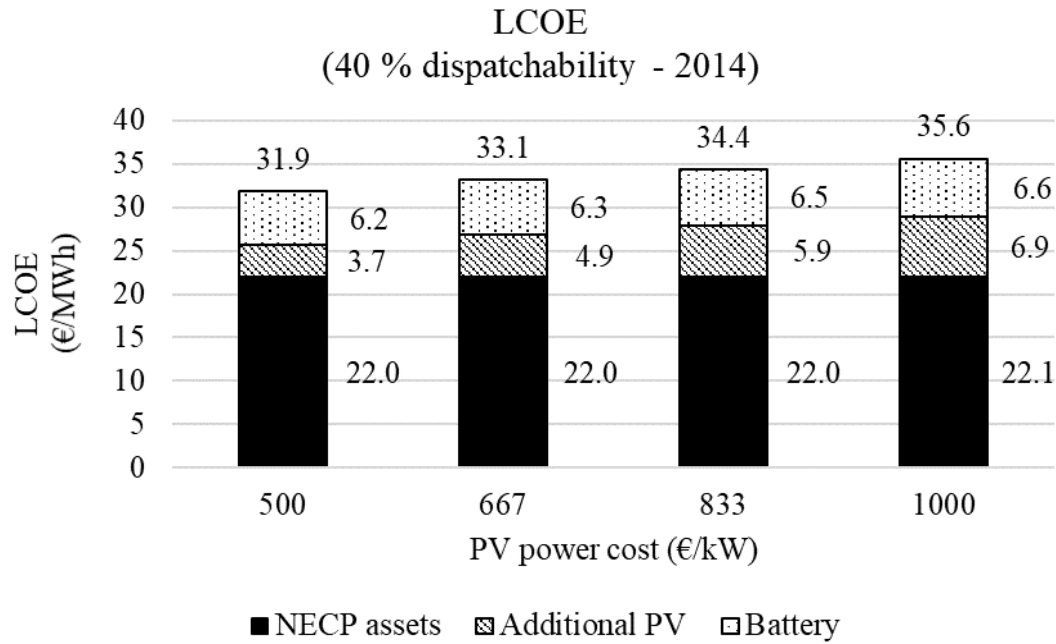

*Figure 10. Sensitivity of overall LCOE to PV power investment costs.*

The increase of the PV investment costs yields two cumulative effects on the LCOE: 1) an increase of the LCOE directly produced by the higher PV costs; 2) indirectly, owing to the increase of PV costs, the optimal configuration of the additional PV power and batteries, required to fill the energy gap, shifts towards a decrease of the PV installed power and an increase of the storage capacity, which also increases the LCOE.

Figure 11 shows the variation of the optimal installed capacity for both technologies, PV and batteries, provided by the optimization procedure.

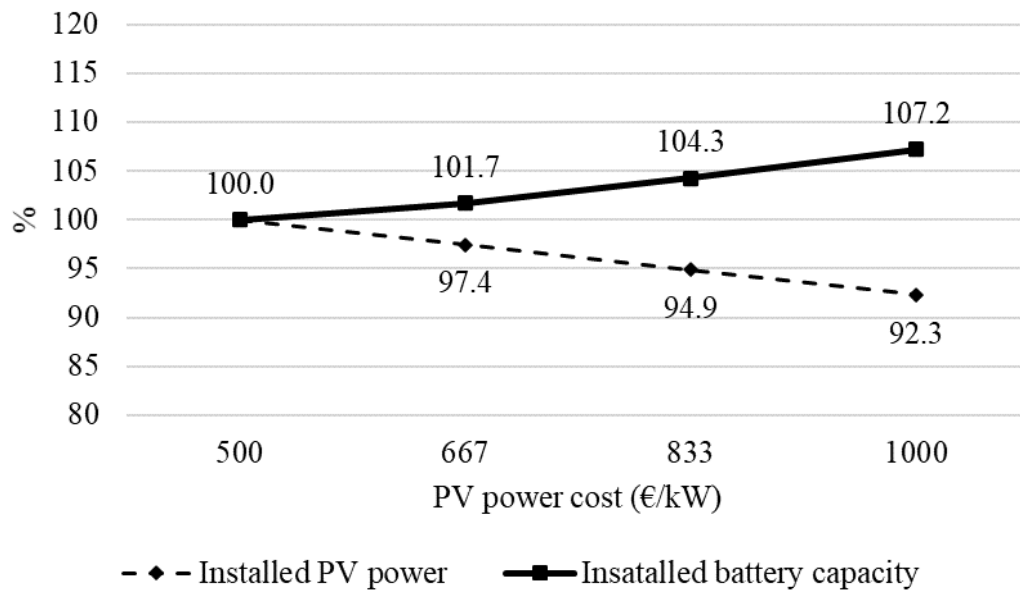

*Figure 11. Installed PV power and storage capacity variation with respect to the base case for increasing investment costs of PV facilities.*

### 3.4.5. Battery cost.

The results obtained above lead to relatively low LCOE values, in general smaller than current wholesale electricity markets in Europe. However, those base case values are obtained with a forecasted battery cost of 100 €/kWh. As mentioned in section 3.1, raw materials scarcity, high demand, etc. may significantly impact the battery storage market prices, and hence the resulting LCOE.

This section compares the LCOE for several values of the battery storage investment costs: 100 €/kWh (base case), 150 €/kWh, 200 €/kWh and 250 €/kWh. Figure 12 shows the results for the year 2014.

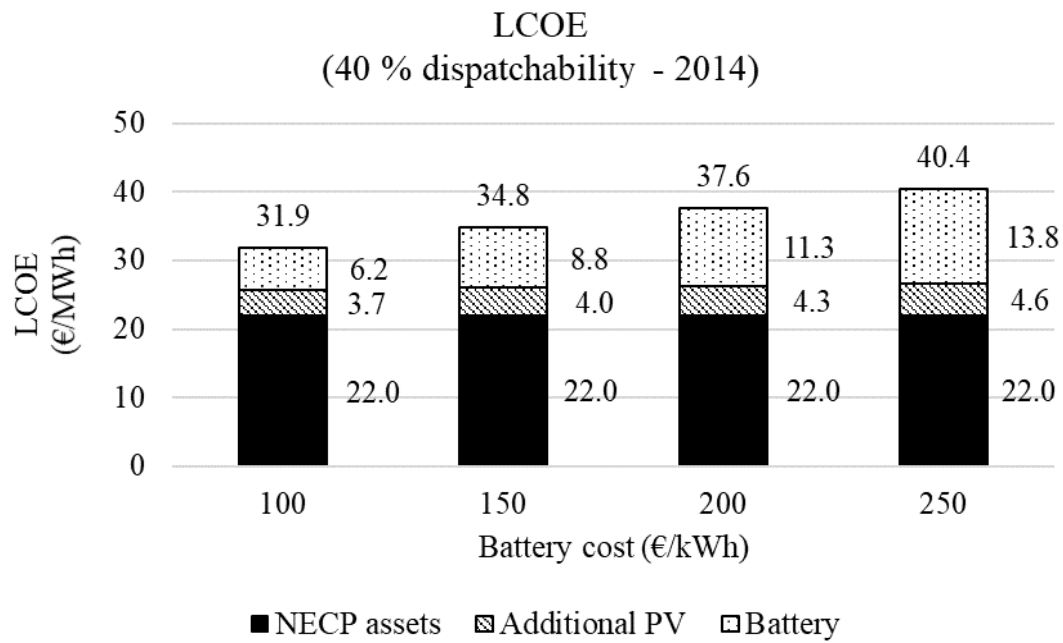

*Figure 12. Sensitivity of overall LCOE to battery storage investment costs.*

By reducing the cost of battery storage, the PV power considered in the optimization algorithm decreases, while at the same time the battery capacity required to match the load increases. In other words, as the relative PV/battery costs increase, the PV installed power is partially replaced by a higher capacity storage system, lowering the total cost.

Figure 13 shows the percentage change of the optimal installed PV power and storage capacity provided by the methodology. As can be seen, the relative increase of PV installed power is roughly symmetrical with the relative decrease of storage capacity.

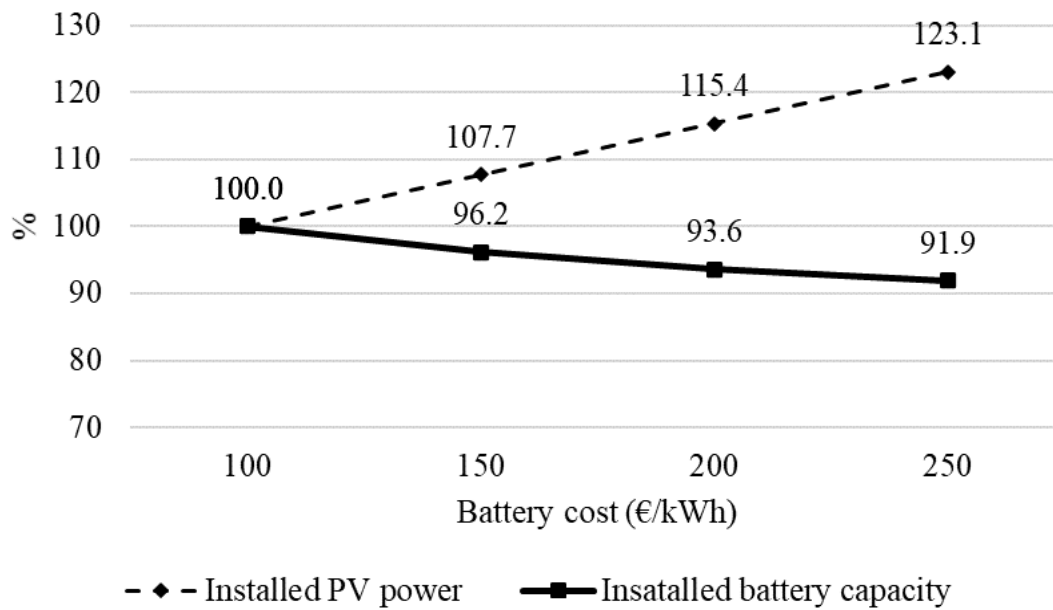

*Figure 13. Installed PV power and storage capacity variation with respect to the base case for increasing investment costs of battery storage*

## 4 Discussion of results.

In the above sections, a future hypothetical scenario is considered for Spain, without nuclear and combined cycles, in which the resulting electricity gap is covered with a least-cost combination of PV plus battery storage, leading to a portfolio that is almost 100% renewable. Such base-case scenario is replicated eleven times, for the generation profiles corresponding to all years of the last decade plus a synthetic year representing the worst-case combination of RES production and demand. Then, for assumed investment costs of PV power and battery storage, the resulting LCOEs are computed for every year. The results obtained in the Spanish case show that, even considering a hydro dispatchability of just 40% (conservative scenario), the system LCOE for the worst-case year would be 38.6 €/MWh, with a minimum value of 31.9 €/MWh in 2014, an exceptional year in terms of demand and generation conditions. Even in the worst case, the overall LCOE would remain below the current market levels, which is good news.

Moreover, a sensitivity assessment is performed with respect to five parameters, classified in three categories:

1. Degree of dispatchability associated with existing RES, particularly hydro generation.
2. Weather-related parameters (distribution of generation in time): coincidence and utilization factors for wind and non-dispatchable hydro generation.
3. Cost-related parameters: battery and PV installation costs (ground vs roof).

Table 10 gathers the computed LCOE elasticities for all the variables considered, as well as their variation ranges.

*Table 10: LCOE elasticity vs. different variables.*

| Parameter | Elasticity LCOE | Range |
|---|---|---|
| Dispachability | -0.04 | 1.13 |
| Wind UF | -0.44 | 0.18 |
| Hydro UF | -0.09 | 1.40 |
| Battery cost | 0.19 | 1.50 |
| PV roof % | 0.17 | 1.00 |

Regarding the first group, it has been observed that, as expected, increasing the level of hydro dispatchability reduces the LCOE, as hydro is acting as a natural storage. The observed elasticity of the LCOE to hydro dispatchability is –0.04, which means that each 10% increase in dispatchability would reduce on average the LCOE by 0.4%. Thus, despite the range of dispatchability considered being large (40%-85%), its impact is not very relevant.

For the second group, the influence of the coincidence factor is not statistically significant (see the high p-values in Figure 7) and is therefore ignored in Table 10. However, the utilization factor correctly explains the variation in annual costs, particularly that of the wind resource. Throughout the period considered, the wind utilization factor fluctuates between 1814 and 2138 hours, leading to LCOEs in the range 38.6-31.9 €/MWh, corresponding to an elasticity of –0.44 (i.e., 4.4% reduction in LCOE for every 10% increase of the wind UF). Note that although the sensitivity is large, the range of variation is rather small (18%). Somewhat the opposite happens with hydroelectricity, showing a much lower elasticity (–0.09) but a much larger range of variation (140%). When both factors are combined, it turns out that the overall impact of the hydro resource variability on the LCOE of the Spanish mix becomes more relevant.

Finally, regarding the costs of batteries and PV installations, both have similar elasticities, close to 0.2. Given that, in recent years, they have experienced cost reductions of nearly 20% per year [55], if this trend continues, the expected LCOE would be reduced annually by 8%.

## 5 Conclusions

The aim of this work was twofold: 1) to establish a systematic methodology to analyze the sensitivity of the results provided by the model developed in [33] to several major factors affecting the system LCOE, such as share of dispatchable generation, weather-related variability (mainly wind and hydro, but also temperature, which affects the demand), and investment costs of batteries and PV facilities (mainly conditioned by the location of the plants: ground or rooftop); 2) to apply the assessment methodology to a big system, such as the Spanish one, throughout a whole decade, including the worst-case combination of generation and demand. In this way, the three research questions formulated in the introduction are addressed.

Regarding the design of a resilient system that could withstand extreme conditions, according to the experience from the last decade, it has been found for the Spanish system that the deployment of PV and battery storage assets would have to be increased by 60% and almost 30%, respectively, with respect to the installed capacities that would result if only the 2019 weather and demand conditions were taken into account [33]. This represents the excess capacity that will be needed to hedge against adverse combinations of demand and non-dispatchable generation, such as those recently observed in the cold events Filomena (Spain) or the rare one in Texas.

As expected, the average cost to serve electricity gets reduced when the share of dispatchable hydro increases, albeit the impact is not significant (0.4% reduction in LCOE for every 10% increase in dispatchability) provided at least 40% of hydro can be shifted in time according to the system needs.

As far as the weather-related characteristics are concerned, both the wind and hydro utilization factors have a large impact on the LCOE. While the former features a narrow range of variability, somewhat the opposite happens with the hydro utilization factor. However, despite this asymmetry, both effects are significant in the Spanish system.

In response to the last research question, also as expected, it has been found that reducing the costs of PV and storage installations will result in a significant reduction in the LCOE (2% reduction of LCOE for every 10% reduction in any of both costs). Given that, in recent years, both PV and storage costs have enjoyed about 20% reductions per year, if this trend continues, it is expected that the LCOE values forecasted in this work get reduced by up to 8% per year.

Considering the above results, it can be concluded that, under the hypotheses and conditions analyzed in this work for the Spanish case, it is possible to design an emission-free electricity system that, taking advantage of existing sustainable assets, meets the long-term needs by providing a reliable supply, at an average cost which is noticeably lower than current wholesale market prices. The methodology devised in this work, could be easily applied to any other system and latitude.

## Annex: calculation algorithm.

The optimization algorithm, fully described in [33], can be summarized as follows:

1) First, the hourly generation profile for each technology and the hourly electric demand is identified for a certain year.

2) Secondly, the dispatchable energy is separated from the regular hydraulic production depending on the dispatchability share selected for the analysis (40%, 55%, 70% or 85%).

3) The third step estimates the minimum additional PV capacity which is necessary to meet the annual energy balance, assuming an unlimited storage system:

$$E_{pv} = D - E \qquad (A)$$

where $E_{pv}$ is the annual PV electricity production, $D$ is the annual demand and $E$ is the annual electricity produced by the assets considered in the NECP.

4) Starting from the minimum PV capacity provided by (A), the required storage capacity is sized by performing an estimation of the battery system capable of meeting the hourly demand during the whole year. The minimum suitable value is the one satisfying the hourly balance equation:

$$\sum_{t=1}^{8760}(E_{total}(t) - D(t) + \Delta E_{bat}(t)) + E_{disp} = 0 \qquad (B)$$

where $t$ is an hourly index, $\mathbf{E_{total}}(t)$ is the total hourly energy production (NECP assets plus PV), $\Delta E_{bat}(t)$ is the hourly charged/discharged energy exchanged by the storage system and $E_{disp}$ is the available dispatchable energy.

With an unlimited storage system, arbitrary amounts of energy can be arbitrarily shifted in time throughout the whole year. However, as the installed storage capacity decreases, a value is reached for which the problem is infeasible for one or more hours. Thus, an iterative process is performed to obtain the minimum feasible value by starting from a sufficiently large storage capacity and then reducing it by small discrete steps.

5) The discharging/charging process consists of obtaining energy from the storage system ($\Delta E_{bat}(t) < 0$) when the hourly balance is negative $E_{total}(t) - D(t) < 0$, until the balance is null or the batteries are empty, and charging the batteries with the energy surplus ($\Delta E_{bat}(t) > 0$) when the hourly balance is positive, $\mathbf{E_{total}}(t) - \mathbf{D}(t) > 0$, until all the surplus energy is depleated or the stored energy reaches 100% of the installed capacity.

6) The state of charge of the system is defined as follows:

$$SoC(t) = \begin{cases} C, & \text{for } SoC(t-1) + \Delta E_{bat}(t) \geq C \\ 0, & \text{for } SoC(t-1) + \Delta E_{bat}(t) = 0 \\ SoC(t-1) + \Delta E_{bat}(t), & \text{otherwise} \end{cases} \quad (D)$$

where $SoC(t)$ is the state of charge at hour $t$, $C$ is the maximum installed capacity and $\Delta E_{bat}(t)$ is the charged/discharged energy derived from the balance requirements, equation (B).

7) When the minimum suitable storage capacity is obtained, the LCOE is estimated for the resulting battery size and the installed power of each RES technology, including both the NECP provisions and the additional PV (utility-scale and/or rooftop, depending on the scenario being considered).

8) Then, the PV capacity is increased by a small discrete amount. With the storage system estimated in the previous run, the energy balance is positive,

$$\sum_{t=1}^{8760} (E_t(t) - D(t) + \Delta E_{bat}(t)) + E_{disp} > 0 \quad (E)$$

So, the storage installed capacity must be reduced to cope with the additional PV assumed, by repeating the iterative procedure of steps 4-7.

9) Once the optimal configuration of PV plus storage is updated, the LCOE is re-estimated accordingly. The process is repeated, with increased values of PV installed power, successively leading to smaller amounts of storage capacity. The iterative procedure stops when the storage capacity cannot be further reduced if the energy balance is to be satisfied at every hour of the year. In this way, all the possible combinations of RES production and storage systems are analysed.

10) The outcome of the procedure is the RES generation portfolio and the associated storage system with lowest LCOE, for a given year.